\documentstyle[prd,aps]{revtex}

\newcommand{\rf}[1]{(\ref{eq:#1})}
\newcommand{\be}{\begin{equation}}
\newcommand{\te}{\end{equation}}

\newcommand{\p}{\partial}
\newcommand{\op}{\partial\!\!\! \slash }

\begin{document}
\draft
\title{Spin and Isospin \\ in\\ Quaternion Quantum Mechanics}
\author{ M. D. Maia\\
Universidade de Bras\'{\i}lia, Instituto de F\'{\i}sica\\ 
Bras\'{\i}lia, DF. 70910-900\\
maia@fis.unb.br}
\twocolumn[
\maketitle
\widetext

\begin{abstract}
The algebraic consistency of spin  and isospin  at the  level of  an unbroken
$SU(2)$ gauge theory suggests  the existence of an additional angular momentum
besides the spin and isospin and  also produces a full quaternionic spinor
operator. The latter  corresponds to  a vector boson in  space-time,
interpreted as  a $SU(2)$ gauge field. The existence of  quaternionic  spinor
fields implies in a quaternionic  Hilbert space and its necessary mathematical
analysis. It is shown how  to  obtain a unique representation of a quaternion
function by a   convergent positive power series.
 
\end{abstract}
\twocolumn]
\narrowtext

\pacs {03.65.Bz, 03.65.Ca,11.15.Kc, 02.30.+g}

\section{Quaternion Quantum Mechanics}  

After six decades, quaternion quantum mechanics is coming out of age. The
earliest known reference on a possible generalization  of  quantum theory  with
respect to the background field  dates back to 1934 with the paper by Jordan et
all \cite{Jordan}.   The use of quaternions properly  was  proposed by Birkoff
and von Neumann in 1936, later  developed by  Finkelstein \cite{Fink}, and more
recently  by Adler and others \cite{Adler}.  

The original motivation  for  quaternion quantum mechanics was  formal:
The propositional calculus implies that it is  possible to represent the pure
states of a quantum  system by rays on a Hilbert defined on any associative
division algebra. This   includes the quaternion algebra as the most  general
case. Contrasting with this, the next  division (but non associative) algebra,
the octonion algebra, has been  always associated with physical arguments,
notably in connection with the  $SU(3)$ gauge symmetry and strong interactions
\cite{Pais,Tiomno,Oliveira,Brumby}.

In its essence,  quaternion quantum mechanics  is  a modification of the
complex quantum theory, in which  the  wave functions  belong to  a Hilbert
space  defined over the quaternion field. As a physical theory, it should prove
to be  effective at some  high energy level, exhibiting  experimental evidence
which would  distinguish it from  the  complex  theory \cite{Peres}.  

As in all Clifford algebras, it  is possible  to represent the quaternion algebra as a tensor
product of two independent  complex algebras. Therefore, in principle we could
write quaternion functions  with complex  components and use two
independent expressions  of complex analyticity.  However this does not mean
that  complex concept of analyticity   extends trivially  to quaternions. 

This paper has two purposes. One of them  is to show  
by a simple argument that the   combined  spin and isospin states
leads to quaternion quantum mechanics, and vice versa, that the 
algebraic consistency of quaternion quantum mechanics necessarily  implies in
the existence of  an intrinsic angular momentum representing  the combined spin
and isospin at the level of  an unbroken $SU(2)$ gauge theory. 
The   second and more formal topic is  a consequence of the first:
The emergence of  a  quaternionic Hilbert space requires
the solution of differential equations involving  functions of a quaternionic
variable, quaternionic Fourier expansions and  quaternionic phase
transformations. 
We will see  that the traditional complex notion of analyticity characterized
by direction independent  derivatives does not necessarily apply, so that 
quaternionic quantum states may exhibit different behavior  along different
directions in space.

\section{Isospin and Quaternions}

The standard textbook   explanation on why  quantum mechanics should be defined
over the complex field is based on the double slit experiment, together
with the  complex phase difference for the  wave functions. 
However, this  can also
be  explained by  a real quantum theory provided a special operator
such that  $J^{2}=1$ and  $J^{T}=-J$ is introduced \cite{Fink}.
Although most people would agree that  this  is equivalent to quantum mechanics
over the complex field, a more definitive argument for the complex algebra
comes from the  spin. The  existence and classification of the spinor
representations of the rotation  subgroup of the Lorentz group demands a
solution of   quadratic algebraic equations for the eigenvalues of  the
invariant operators.  This can  be  guaranteed only within  the complex field.
The bottom line is,  complex quantum mechanics is a requirement
of the spin and its  spinor structure  \cite{Chevalley}.

Following a similar argument  we may  convince ourselves
that  quaternion quantum mechanics  is  an algebraic requirement of the spin
together with the isospin and the associated spinor structures.  In fact, while  the
spin is  associated with the spinor representations of the  $SO(3)$ subgroup
of the  Lorentz  group,  the
isospin is given by a representation of  the gauge group $SU(2)$.
 Actually the  two groups are isomorphic and they have  
equivalent  representations given by the Pauli matrices (they are
identical representations, distinguished only by
different notations), acting  on independent  spinor spaces. 

In the case of the unbroken $SU(2)$ gauge theory, 
spin and isospin   are  present in a  combined symmetry scheme, so that
 the total  spinor  space is  a direct sum of the  spinor  space
 and  the isospinor spaces:  $K=I\oplus J $. 
The  spinor  space is represented as
 a complex plane (a Gauss plane) generated by the basis
\[
\mbox{I= (spinor space) }: \left\{ 
1= \left( \begin{array}{l} 1\\ 0  \end{array} \right), 
i= \left( \begin{array}{l} 0\\ 1  \end{array} \right) 
\right\}
\]
and the  isospinor space  represented by  another complex
plane generated by 
\[
\mbox{J=(isopinor space) }: \left\{ 
1= \left( \begin{array}{l} 1\\ 0  \end{array} \right), 
j= \left( \begin{array}{l} 0\\ 1  \end{array} \right) 
\right\}
\]
Thus, $I$  and  $J$ can be taken as two  independent Gauss planes  sharing the
same  real unit  $1$, but with  different  and independent imaginary units $i$ and $j$
respectively.
As long as the isospin symmetry and the spinor
representation of the Lorentz group remain combined,   the corresponding
angular momenta  add up to generate a total  angular momentum  
represented in the direct sum of the  two spinor spaces.
On the other hand, it has been  argued that  when this combined symmetry  is broken
the $SU(2)$  degree of freedom reappears  as  a spin  degree of freedom
\cite{Jackiw}.  Following these ideas a fermionic state is derived from
a  magnetic monopole model in four dimensions associated with an $SU(2)$
soliton \cite{Hooft,Vachaspati,Singleton,Emch}. 

The  question we address  here  concerns   with the algebraic consistency  of
the  combined spinor-isospinor symmetry as  it remains unbroken. In this case,
we end up with the total spinor space generated by the  hypercomplex basis
$\{ 1,i,j\} $. According to Hamilton this direct sum is 
 algebraically   consistent only  if  a third  imaginary unit $k$ such that
 $k=ij$  is introduced. That is,  we can only close the algebra if
 a   third complex plane
\[\mbox{K=( new spinor space)} :\left\{ 
1= \left( \begin{array}{l} 1\\ 0  \end{array} \right), 
k= \left( \begin{array}{l} 0\\ 1  \end{array} \right) 
\right\}
\]
is introduced. We conclude that  an additional spin-half  field  should also be
present. This new  spinor may be the 
generator of the Jackiw-Rebbi  spin degree of freedom after the symmetry is
broken. However, this is not all. When the combined symmetry remains
unbroken  the  three spinors  produce a  full quaternion algebra with basis
$\{1,i,j,k\}$, such that  its group  of automorphisms
carry  the combined symmetry. If  we  add to the  space of these  quaternion
wave functions a   Hilbert product  compatible with the quaternion algebra we
obtain a  quaternionic  Hilbert space. 
This  space should reduce to the  usual  Hilbert space of 
 complex quantum mechanics with separate spin, isospin plus one extra spinor
degree of freedom at the  level of the combined symmetry breaking. 

In conclusion,  quaternion quantum mechanics  appears  as consistent condition
of the combined spin and isospin symmetries.
A possible relation between quaternions and the  isotopic spin   was suggested
by C. N. Yang  \cite{Fink} and  by E. J. Schremp  \cite{Schremp}.
However, their  basic arguments are   distinct from the ones based on  the
combination of symmetries. 

When the two  spinor spaces are taken together, they give way to a  quaternionic
spin operator given by a  $2\times 2$ 
matrix representation of the quaternion algebra given by the  Pauli matrices
\begin{eqnarray*}
\sigma^{0}=\left(
\begin{array}{cc}
1 & 0\\
0 & 1
\end{array}
\right),\;\;
\sigma^{1}=\left(
\begin{array}{cc}
0 & 1\\
1 & 0
\end{array}
\right),
\\
\sigma^{2}=\left(
\begin{array}{cc}
0 & -i\\
i & 0
\end{array}
\right),\;\;
\sigma^{3}=\left(
\begin{array}{cc}
1 & 0\\
0 & -1
\end{array}
\right)
\end{eqnarray*}
This  has  a  one to one  correspondence  with  the  quaternion
algebra\footnote{ Greek indices are space-time indices  and they  run from 0 to 3. Small case
Latin indices run from 1 to 3. Capital Latin indices are spinor indices running
from 1 to 2. The quaternion multiplication table  is  taken to be
\[
e_{i}e_{j} = -\delta_{ij} +\sum\epsilon_{ijk}e_{k},\;\;\;
e_{i}e_{0} = e_{0}e_{i} =e_{i}
\]
The conjugate of  a quaternion $X$ is  $\bar{X}$, with
$\bar{e}_{i}=-e_{i},\;\; \bar{e}_{0}=e_{0}$. The quaternion norm is
$|X|^{2}=\sum (X^{\alpha})^{2}$ and the inverse of  $X$ is
$X^{-1}=\bar{X}/ |X| $.}
whose elements are the quaternions

\be
\Psi  =  \sum_{\alpha=0}^{3} \Psi_{\alpha}\sigma^{\alpha}
=\left(
\begin{array}{cc}
\Psi_{0}+\Psi_{3} & \Psi_{1} - i\Psi_{2}\\
\Psi_{1}+ i\Psi_{2} & \Psi_{0}-\Psi_{3} 
\end{array}
\right),  \label{eq:Q}
\te

As it is well known, the above matrix  corresponds to  a
vector field in space-time, associated to  a pair  of  two component
spinors $\Psi^{\alpha} =\sigma^{\alpha}_{AB}\xi^{A}{\bar{\zeta}}^{B}$.
A possible interpretation for this pair of  spinors  is 
given by the  spin and isospin states in a  $SU(2)$ model
\cite{Hooft,Vachaspati,Singleton,Emch}.   

Since the inner automorphism of the quaternion algebra
correspond to the isometries of space-time, the   existence of  a  combined
spin-isospin structure also have implications on the  classical notion of derivative
of a function. In ordinary complex analysis 
the derivative of  a  function does not depend on the  direction in the complex
plane along which the limit is taken. This  has to be so because the  complex
plane is  generated by only one real and one imaginary direction, leading to
the Cauchy-Riemann conditions for analyticity. 
On the other hand, in the  the quaternionic case there are  three imaginary
directions generating a space that is isomorphic to $I\!\!R^{3}$, 
where in principle there is no reason for  the  derivatives to be all equal.  
This means that  the properties of differentiable equations
involving quaternion functions of a quaternionic variable  does not necessarily
coincide with  those of  ordinary complex quantum theory. Fortunately this can
be examined  through the methods of  classical analysis.

\section{Analysis of Quaternion Fields}

The earliest  known study on the analysis of  quaternion functions 
using the  same  concepts of complex  analysis
was  made by Fueter in 1932, finding  very restrictive generalizations of
the Cauchy-Riemann  conditions \cite{Fueter}. Some alternative
 criteria for defining   quaternion analyticity   have been suggested
\cite{Ketchum,Ferraro,Maia,Nash,Gursey,Khaled}, but to date  there is not a
consensus on what is meant by  quaternion analyticity. 
To understand the nature of the difficulties we need to start from the basic
principles. 
Denoting a  generic  quaternion function  by  $ f(X)=\sum
U_{\alpha}(X)e^{\alpha}$, and  $\Delta f= [f(X+\Delta X) -f(X)]$ we may define
its left  derivative  as 
\[
f'(X) = \mbox{lim}_{\Delta X\rightarrow 0}\delta f(X) (\Delta X)^{-1},
\]
and the right derivative as 
\[
'f(X) = \mbox{lim}_{\Delta X\rightarrow 0} (\Delta X)^{-1}\Delta f(X)
\]
where the  limits are  taken  with  $|\Delta X|\rightarrow 0$ 
along  the direction of the four-vector $\Delta X$  which depends on the
3-dimensional vector $\vec{\Delta X}$. 

To compare with the complex case, 
we may use  the exponential form  of a  quaternion: 
A  vector  of  $I\!\!R^{3}$,  $\xi =\sum X_{i}e^{i}$ 
associate a quaternion $\xi$ with norm 
$\vert \xi\vert^{2}=\xi\bar{\xi}=\sum X_{i}^{2}$  and  square $\xi^{2}
=-\xi\xi=-\sum X_{i}^{2}$. 
Defining the unit  quaternion (iota)
$\iota ={\xi}/{\sqrt{\xi\bar{\xi}}} $  such that  $\iota^{2}=-1$,
 we may construct a Gauss  plane  generated by $\iota$ and the quaternion unit
 $e^{0}$. In this plane a   quaternion $X=X_{0}e^{0}+\sum X_{i}e^{i}$ can be
expressed  as $ X=\vert X\vert (cos\gamma +\iota\sin\gamma) $,
with  $tan\; \gamma = {X_{0}}/{\sqrt{\sum X_{i}^{2}}}$.
Therefore, if   we define the  quaternion
exponential  as
\[
exp(\iota\gamma)= e^{\iota\gamma} =cos\gamma +\iota sin\gamma, 
\]
then we may express
$\Delta X=\vert\Delta X\vert e^{\iota\gamma}\;\;\mbox{and}\;\;\Delta
X^{-1}={ e^{-\iota\gamma}}/{\vert\Delta X\vert}$, where the three dimensional
direction is included in the definition of  $\iota$.

Contrasting with the complex case,   we cannot neglect the phase factor
$e^{\iota \gamma}$ during  the limiting process because we have a functions
depending on three variables. The independence of the 
limit with the phase is a privilege (or rather a limitation) of  complex
theory. Furthermore,  here we have the added complication that the left and
right derivatives do not necessarily coincide.  
To see the consequences of this, consider the derivatives of  a quaternion
function $f(X)$ along a  fixed direction  $\Delta X =\Delta X_{\beta} e^{\beta}
$ (no sum), indicated by the  index within parenthesis
\begin{eqnarray}
f'(X)_{(\beta)}   =  \frac{\partial U_{0}}{\partial X_{\beta}}e^{0}
(e^{\beta})^{-1} + 
\sum_{i}\frac{\partial U_{i}}{\partial X_{\beta}}e^{i}(e^{\beta})^{-1}\\
'f(X)_{(\beta)}   =  \frac{\partial U_{0}}{\partial X_{\beta}}
(e^{\beta})^{-1}e^{0} + 
\sum_{i}\frac{\partial U_{i}}{\partial X_{\beta}}(e^{\beta})^{-1}e^{i}
\end{eqnarray}
From  direct calculation  we find  that
\begin{eqnarray*}
f'(X)_{(0)}& =& \,  'f(X)_{(0)}\\
 f'(X)_{(j)}&=& \,  'f(X)_{(j)} 
-2\sum_{i,k}\epsilon^{ijk}\frac{\partial U_{i}}{\partial X_{j}} e^{k}
\end{eqnarray*}
Imposing that these derivatives  are selectively  equal,
four basic  classes of complex-like analytic  functions are obtained 
\begin{description}
\item[ Class A)] 
Right analytic functions
\be
\left.
\begin{array}{ll}
\! f'(X)_{(0)} \! =   f'(X)_{(i)}\\
\! f'(X)_{(i)} \! =  f'(X)_{(j)}
\end{array}
\right \}
  \Rightarrow
          \left\{ 
         \begin{array}{ll}
\frac{\partial U_{\alpha}}{\partial X_{\alpha}} = \frac{\partial U_{\beta}}{\partial X_{\beta}}\\
\frac{\partial U_{i}}{\partial X_{0}} = -\frac{\partial U_{0}}{\partial X_{i}}\\
\frac{\partial U_{i}}{\partial X_{j}} = \sum\epsilon^{ijk}\frac{\partial U_{k}}{\partial X_{0}}
          \end{array}
		  \right.
\label{eq:A}
\te
\item[Class B)]  Left  analytic functions
\be
\left.
\begin{array}{ll}
\! 'f(X)_{(0)}\!  =  \,  'f(X)_{(i)}\\
\! 'f(X)_{(i)}\!  =\,   'f(X)_{(j)} 
\end{array}
\right\}
 \Rightarrow
          \left\{
		  \begin{array}{ll}
\frac{\partial U_{\alpha}}{\partial X_{\alpha}}=\frac{\partial U_{\beta}}{\partial X_{\beta}}\\
\frac{\partial U_{i}}{\partial X_{0}}=-\frac{\partial U_{0}}{\partial X_{i}}\\
\frac{\partial U_{i}}{\partial X_{j}}=-\sum\epsilon^{ijk}\frac{\partial U_{k}}{\partial X_{0}}
          \end{array}\right.
\label{eq:B}
\te
\item[Class C)]  Left-right analytic functions
\be
f'(X)_{(\alpha)}=  \,  'f(X)_{(\alpha)}
  \Rightarrow
            \left\{
			\begin{array}{ll}
 \frac{\partial U_{\alpha}}{\partial X_{\alpha}}=\frac{\partial
U_{\beta}}{\partial X_{\beta}}\\ 
 \frac{\partial U_{i}}{\partial X_{j}}=-\frac{\partial U_{j}}{\partial X_{i}}\\
 \frac{\partial U_{i}}{\partial X_{0}}=-\frac{\partial U_{0}}{\partial X_{i}}\;\;\;\;
          \end{array}
		  \right.
\label{eq:C}
\te
\item[Class D)]  The total analytic functions
\be
\left.
\begin{array}{lll}
\! f'(X)_{(\alpha)}\!  =\!  f'(X)_{(\beta)}\\
\! 'f(X)_{(\beta)}\!  =\! \, 'f(X)_{(\alpha)}\\
\! 'f(X)_{(\alpha)}\!=\! f'(X)_{(\alpha)}
\end{array}
\right\}
  \Rightarrow
            \left\{
			\begin{array}{ll}
 \frac{\partial U_{\alpha}}{\partial X_{\alpha}}=\frac{\partial U_{\beta}}{\partial X_{\beta}}\\
 \frac{\partial U_{\alpha}}{\partial X_{\beta}}= 0\;\;  \alpha\neq \beta
          \end{array}
		  \right.
\label{eq:D}
\te

\end{description}
As we can see, these  conditions are very restrictive, specially when we
consider the  applications to   quantum mechanics, suggesting
the adoption of different  criteria for analyticity.

\section{ Harmonicity}

Consider the operator 
$\op =\sum  e^{\alpha}\p_{\alpha}=\sum e^{\alpha}{\p}/{ \partial X_{\alpha}}$
acting on the  right and on the left  of a function $f(X)$
\begin{eqnarray}
\op f(X) & = & (\frac{ \partial U_{0}}{ \partial X_{0}} +
\sum \frac{\partial U_{i}}{\partial X_{0}}e^{i})\nonumber\\
        & +  & \sum[\frac{\partial  U_{0}}{\partial X_{i}}e^{i}
- \sum \frac{\partial U_{i}}{\partial X_{j}}(\delta^{ij}
-\epsilon^{ijk}e^{k})] \nonumber\\
       &  =  &    'f(X)_{(0)} + \sum_{j}\, 'f(X)_{(j)}\vspace{3mm}\\
f(x)\op & =  & (\frac{\partial U_{0}}{\partial X_{0}} +\sum \frac{\partial
U_{i}}{\partial X_{0}}e^{i}) \nonumber\\
       &  + & \sum[\frac{\partial  U_{0}}{\partial X_{i}}e_{i} 
-\sum \frac{\partial U_{i}}{\partial X_{j}
}(\delta^{ij}+ \epsilon^{ijk}e^{k})]\nonumber\\
       & =  & f'(X)_{(0)} - \sum_{j} f'(X)_{(j)} 
\end{eqnarray}
Using these results   four new classes of  quaternion functions can be defined:
\begin{description}
\item[Class E)]
The   functions such that
\be
\op f(X)= f(X)\op   \Rightarrow
          \left\{
  \begin{array}{ll}
\frac{\partial U_{i}}{\partial X_{j}}= \frac{\partial U_{j}}{\partial X_{i}}
   \end{array}
		\right.
\label{eq:E}
\te
\item[Class F)]  The left  harmonic functions
\be
\op f(X)\!=\! 0  \Rightarrow
          \left\{
		  \begin{array}{ll}
 \frac{\partial U_{0}}{\partial X_{0}}=\sum_{i} \frac{\partial U_{i}}{\partial X_{i}}\\
 \frac{\partial U_{k}}{\partial X_{0}} +\frac{\partial U_{0}}{\partial
X_{k}}\!=\! \sum_{ij}\epsilon^{ijk}\frac{\partial U_{i}}{\partial X_{j}}
         \end{array}\right.
\label{eq:F}
\te
\item[Class G)]  The right  harmonic functions 
\be
f(X)\op\!=0\! \Rightarrow\!
          \left\{
		  \begin{array}{ll}
 \frac{\partial U_{0}}{\partial X_{0}}=\sum_{i} \frac{\partial U_{i}}{\partial X_{i}}\\
 \frac{\partial U_{k}}{\partial X_{0}} +\frac{\partial U_{0}}{\partial
X_{k}}\!=\!-\!\sum_{ij}\epsilon^{ijk}\frac{\partial U_{i}}{\partial X_{j}}
   \;\;\; \end{array}\right.
\label{eq:G}
\te
\item[Class H)] The left and right harmonic functions

\be
\op f(X)\!=\! 0\;\mbox{and}\; f(X)\op\! =\!0\!  \Rightarrow
          \left\{
  \begin{array}{ll}
\frac{\partial U_{0}}{\partial X_{0}}=\! \sum_{i}\frac{\partial U_{i}}{\partial X_{i}}\\ 
\frac{\partial U_{i}}{\partial X_{0}}= \! -\!\frac{\partial U_{0}}{\partial X_{i}}\\
\frac{\partial U_{i}}{\partial X_{j}}=   \frac{\partial U_{j}}{\partial X_{i}}\;\;\;\;\;
   \end{array}
		\right. \label{eq:H}
\te
\end{description}
Notice that for the classes  F, G and  H we have
\[
\sum\delta^{ij}\frac{\partial^{2} U_{0}}{\partial X_{i}\partial
X_{j}}+\frac{\partial^{2} U_{0}}{\partial X_{0}^{2}}=\Box^{2} U_{0}=0
\]
where   $\Box^{2}=\op\bar{\op}$. Similarly,  $\Box^{2}U_{k}=0$,
so that those classes  describe harmonic functions in the sense that
$\Box^{2} f(X)=0$.

A non trivial  example of class H quaternion function is   given by a  
instantons field  expressed in terms of  quaternions \cite{Atiah}. The
connection  of  an anti  self dual $SU(2)$ gauge field is given by the  form
\begin{equation}
\omega =\sum_{\alpha}A_{\alpha(X)}dx^{\alpha} \label{eq:INSTANTON}
\end{equation}
where   $A_{0}=\sum U_{k}e^{k}$ and  $A_{k}=U_{0}e^{k}
-\epsilon_{ijk}U_{i}e^{j} $ and  where 
\[
U_{0}=  \frac{\frac{1}{2} X_{0}}{1+|X|^{2}},\;\;\;
U_{i}=\frac{-\frac{1}{2}X_{i}}{1+|X|^{2} } 
\]
are   the components of the  quaternion function  $f(X)=U_{\alpha}e^{\alpha}$.
We can  easily see that   $f(X)$  satisfy the  conditions  \rf{H}
in the  region of  space-time  defined by  $\sum X_{i}^{2}=-2X_{0}$. In fact, this
is  a particular case  of  a  wider class of quaternion functions with
components  $U_{\alpha} =g_{\alpha}(X)/(1 +|X|^{2})$, where $g_{\alpha}(X)$ are
some real  functions. The case of instantons correspond to the choice
$g_{0}=\frac{-1}{2} 
\sum X_{i}^{2}$ and $g_{i} =\frac{\partial g_{0}}{\partial X_{i}}$.
It is also interesting  to notice that  the anti instantons do not  belong to
the same class of analyticity as the instantons.

\section{Integral Theorems}
Given  a quaternion function $f(X)$ defined on a  orientable  3-dimensional
hypersurface 
$S$ with, unit normal vector $\eta$ we may define two   integrals
\[
\int_{S} f(X) dS_{\eta},\;\,\mbox{and}\,\;\int_{S}  dS_{\eta} f(X)  
\]
where   $dS_{\eta}=\sum dS_{i}e^{i}$  denotes the quaternion hypersurface
element with components
\begin{eqnarray*}
dS_{0}=dX_{1}dX_{2}dX_{3},\;\; dS_{1}=dX_{0}dX_{2}dX_{3},\\
dS_{2}=dX_{0}dX_{1}dX_{3}, \;\; dS_{3}=dX_{0}dX_{1}dX_{2}.
\end{eqnarray*}
 On the other hand,
denoting by  $dv=dX_{0}dX_{1}dX_{2}dX_{3}$  the 4-dimensional  
volume element in a region $\Omega$ bounded by $S$, we  obtain
after integrating  in  one of the variables we obtain 
\begin{eqnarray*}
& &\int_{\Omega} \op f(X)dv= 
\int_{\Omega} e^{\alpha}\partial_{\alpha} e^{\beta}U_{\beta} dv =\\
&&\int_{\Omega}[(\p_{0}U_{0}\!-\!\sum_{i}\p_{i}U_{i}) \!+\!
\sum_{i} (\p_{0}U_{i}+\p_{i}U_{0}) e^{i} \! +\!\epsilon^{ijk}\p_{i}U_{j}e^{k}
] dv
\end{eqnarray*}
and noting that
\begin{eqnarray*}
\int_{\Omega}\p_{0}U_{0}dv =\int_{S}U_{0}dS_{0},\,
&&\int_{\Omega}\p_{0}U_{i}dv = \int_{S}U_{i}dS_{0},\\
\int_{\Omega}\p_{i}U_{0}dv =\int_{S}U_{0}dS_{i},\;\;
&&\int_{\Omega}\p_{i}U_{j}dv =\int_{S}U_{j}dS_{i}
\end{eqnarray*}
it follows   that
\begin{eqnarray*}
\int_{\Omega} \op f(X)dv &=& 
 \int_{S}
 [( U_{0}dS_{0}-\sum\delta^{ij} U_{i}dS_{j})e^{0}\\
 & +&\sum(U_{i}dS_{0}
+U_{0}dS_{i} )e^{i} -\sum \epsilon^{ijk} U_{i}dS_{j}e^{k} ]
\end{eqnarray*}
It is  a simple matter to see that  this is  exactly the same expression
of the  surface integral 
\[\int_{S}dS_{\eta}f(X)=\sum \int_{S} U_{\alpha} dS_{\beta} e^{\beta}e^{\alpha}
\]
Therefore,  we obtain the result
\be
\int_{\Omega} \op f(X)dv=\int_{S} dS_{\eta} f(X)  \label{eq:RDIV} 
\te
and similarly  we obtain  for the left hypersurface integral
\be
\int_{\Omega}f(X)\op dv=\int_{S} f(X)dS_{\eta}\label{eq:LDIV}
\te
The  above integrals  hold  for  any of the  previously defined classes of
functions and  they  difference is 
\[
\sum\epsilon^{ijk}e^{k}\!\!\int_{S}(U_{i}dS_{j}\!-\! U_{j}dS_{i})\!=\!
\!-\!\sum\epsilon^{ijk}e^{k}\!\!
\int_{\Omega}(\frac{\partial U_{i}}{\partial X_{j}} + \frac{\partial
U_{j}}{\partial X_{i}})dv 
\]  
which  vanish  on account of  Green's theorem in the $(i,j)$ plane.
The following result extends the  first Cauchy's Theorem for  quaternion
functions: 

{\em If $f(X)$ is  of  H class  in the interior of  a region $\Omega$
bounded by a hypersurface $S$ then }
\be
\int_{S} f(X) dS_{\eta} =\int_{S}dS_{\eta} f(X)=0 \label{eq:CAUCHY1}
\te
This follows immediately from eqns. \rf{RDIV}, \rf{LDIV} and the condition
for  a class $H$ function \rf{H} where $\op f(X)= 0$ and $f(X)\op =0$. 

The second  Cauchy's theorem is  also true only  for  class H functions

{\em If $f(X)$  satisfy the conditions of class $H$, in a  region  bounded by  a
simple closed 3-dimensional hypersurface $S$,  then for  $P \in S $,}
\be
f(P)=\frac{1}{\pi^{2}}\int_{S} f(X)(X-P)^{-3}dS_{\eta} \label{eq:CAUCHY2}
\te
In fact, the integrand  does not satisfy the  class  $H$ conditions in $\Omega$
as it is not  defined   at $P$ and consequently
 the previous theorem does not apply. However this  point may be isolated by a
  sphere with  surface  $S_{0}$ with center at $P$ and radius  $\epsilon$ 
such that it is  completely inside  $\Omega$.  
Applying the previous theorem in the region bounded by $S$  and  $S_{0}$
 we obtain
\[
\int_{S} f(X)(X-P)^{-3}dS_{\eta} +\int_{S_{0}}f(X)(X-P)^{-3}dS_{\eta} =0
\]
Now, the primary  condition for  a function belonging to  class  E through H is
that  its components are  regular so that  we may  calculate their Taylor
series   around  $P$: 
$
U_{\alpha}(X) =  U_{\alpha}(P) +\epsilon^{\beta}\frac{\partial
U_{\alpha}}{\partial x^{\beta}}\rfloor_{P} +\cdots .
$
Using  this  expansion  in the   integral over  $S_{0}$ and taking the limit
 $\epsilon\rightarrow 0$,  it follows that
\be
f(P)\! =\!\left( \int_{S} f(X)(X\!-\! P)^{-3}dS_{\eta}\right) \left(
\int_{S_{0}} (X\!-\! P)^{-3}dS_{\eta}\right)^{-1} \label{eq:fp}
\te
In order to calculate the integral over the sphere it is convenient to use four
dimensional spherical coordinates  
$(r,\theta, \phi,\gamma) $,  such that
 $X_{0}=r sin\gamma $,  $X_{1}=r cos\gamma\, sin\theta\,cos\phi $,
 $X_{2}=r cos\gamma\, sin\theta\,sin\phi $ and  $X_{3}=r cos\gamma\, cos\theta$
 where  $\theta\in(0,\pi)$, $ \phi\in(0,2\pi)$, $ \gamma\in(-\pi/2 , \pi/2)$.
 With this, the  coordinates  $X_{0},X_{1},X_{2},X_{3}$ correspond to the
coordinates of a space-time point with quaternion norm $|X|^{2}=r^{2}$, while
$\gamma $  span values from the past to the future.
Then  the volume element  is  $dv=J dr d\theta d\phi d\gamma$ where
$J=-\epsilon^{3}cos^{2}\gamma\sin\theta$  is the Jacobian determinant. 
Using the polar form, the  unit normal  to the sphere centered at $P$ can be 
written as  $\eta=e^{\iota\gamma}$ and $X-P =\epsilon e^{\iota\gamma}=\epsilon
\eta$, so that
\[
\int_{S_{0}} (X-P)^{-3}dS_{\eta}=\int_{S_{0}}
e^{-2\iota\gamma}sin^{2}\gamma sin\theta\, d\theta\, d\phi\, d\gamma=\pi^{2}
\] 
After replacing  in  \rf{fp} we obtain  the result \rf{CAUCHY2}.

Notice that  the  power  $(-3)$  in \rf{CAUCHY2} is not accidental
as it is  the right  power  required to  cancel  the Jacobian
determinant as  $\epsilon\rightarrow 0$.

\section{Power Series}
To conclude, consider  the particular  function $f(X)=(1-X)^{-3}$, with
$\vert X\vert<1$.
It is  a simple matter to see that  it can  be expanded as 
\begin{equation}
(1\!-\! X)^{-3}\!=\!\sum_{1}^{\infty}\frac{n(n+1)}{2}X^{n-1}\! =\!
\sum_{m=0}^{\infty}\frac{(m\!+\!1)(m\!+\!2)}{2}X^{m}\label{eq:example}
\end{equation} 
Using this  particular  case we  may   prove  the following general  result
for quaternion functions: 

{\em Let $f(X)$ be  of  class H inside a region $\Omega$ bounded by a surface
$S$. Then for all $X$ inside  $\Omega$   there are  coefficients  $a_{n}$ such
that }
\be
f(X) =\sum_{0}^{\infty}  a_{n}(X-Q)^{n}  \label{eq:TH3}
\te
The proof is   a straightforward adaptation from the  similar complex theorem.
If  $S_{0}$ is the largest sphere in $\Omega$ centered at $Q$,
the  integral  \rf{CAUCHY2}  for  a point  $P=X$ inside $\Omega$ gives
\[
f(X)\! =\! \frac{1}{\pi^{2}}\int_{S}\!
f(X')(X'\!-\! Q)^{-3}[1\!-\! (X'\!-\!Q)^{-1}(X\!-\!Q)]^{-3}dS'_{\eta} 
\]
Assuming that     $|X-Q| < |X'-Q|$ and using  \rf{example}, the integrand is
equivalent to 
\begin{eqnarray*}
&&[1-(X'-Q)^{-1}(X-Q)]^{-3}\\
&=&\sum_{0}^{m=\infty}\frac{(m+1)(m+2)}{2}(X'-Q)^{-m}(X-Q)^{m} 
\end{eqnarray*}
so that 
\begin{eqnarray}
&&f(X)=\frac{1}{\pi^{2}}\sum_{m=0}^{\infty}\frac{(m+1)(m+2)}{2}\times\nonumber\\
&&\times\int_{S_{0}}f(X')(X'-Q)^{-3-m} (X-Q)^{m} dS'_{\eta} \label{eq:FP}
\end{eqnarray} 
Now  we may  write
$(X-Q)^{m}=\epsilon^{m}e^{m\iota\gamma}$ and    $dS'_{\eta}=e^{\iota\gamma}dS' $, 
and it follows that 
\[
(X\!-\!Q)^{m}dS'_{\eta}\!-\!dS'_{\eta}(X\!-\!Q)^{m}=\epsilon dS'(e^{m\iota\gamma}
e^{\iota\gamma}\!-\!e^{\iota\gamma}e^{m\iota\gamma})\! =\! 0 
\]
Therefore  \rf{FP}  is  equivalent to
\begin{eqnarray*}
&&f(X)=\frac{1}{\pi^{2}}\sum_{m=0}^{\infty}\frac{(m+1)(m+2)}{2}\times\\
&&\times \int_{S_{0}}f(X')(X'-Q)^{-3-m}dS_{\eta}\, (X-Q)^{m} 
\end{eqnarray*}
or, after defining  the coefficients 
\be
a_{m}\!=\! \frac{1}{\pi^{2}}\frac{(m\!+\!1)(m\!+\! 2)}{2}
\int_{S_{0}}f(X')(X'\!-\! Q)^{-3-m} dS'_{\eta} \label{eq:COE1}
\te
we obtain   \rf{TH3}.

This important result shows that  class $H$  functions can be expressed as  a
convergent positive power  series.
Therefore  the  class H or the latter property could be taken to represent a  class of
analyticity for  quaternion  functions, in the same sense  of the real and
complex analyticity. However,  unlike the complex case from \rf{H} we see that
 their  derivatives  depend on the direction in which the limit is taken.

\section{Discussion}

At the level  of an  unbroken $SU(2)$ gauge  theory
the algebraic  properties of the spinors  predicts  a combined spin-isospin
angular  momentum here called the k-spin (after  i-spin for complex and j-spin
for 
isospin). The  three resulting  spinors  give way to  a full quaternionic
spinor  operator obtained from the  linear  combination of the Pauli matrices
in a specific representation. In this way,  we conclude  that
quaternion quantum mechanics may be effective at the level of the combined
spinor symmetry. The  quaternionic  spinor operator  naturally associates a
vector in space-time whose physical interpretation  depends on the   $SU(2)$
model of gauge field  considered. We have  suggested the 
't Hooft-Poliakov monopole  as  a possible  interpretation of that  vector
field.

The emergence of quaternionic  spinor fields  requires  a
proper analysis of   quaternion functions of  a quaternionic variable.
We have  shown that  there is  a class of  analytic  quaternion functions
 which can  be represented by  a   positive power series, a
property which is shared with  the other two associative division algebras (the
real and complex functions). Outside the conditions for  class H
the  the  power expansions would also  have negative powers  and  the associated
poles as points in space-time and their corresponding  residues \cite{Maia}.

The harmonic property implicit in class H implies in the possibility that the
quaternion  quantum fields and states can  be  represented in terms of
quaternionic Fourier  expansions, something that is  required to represent the
quaternion wave  packets. 
As it has been noted, quaternion analyticity does not
imply that the  derivatives are  independent of direction in space
and in this respect   complex analysis and the
corresponding quantum theory may be considered to be somewhat limited as compared  with
quaternion analysis. The direction dependent property should be detectable at 
the level of the combined symmetry
 
It is  conceivable that the characterization of analyticity either by class H
or  more generally by positive power series expansions will
 not hold at  higher energy levels,
where the  wave functions are subjected to  fast variation at  a sort time. In
this case, the best we may hope  that these functions remain
differentiable and  any appeal to  analyticity in the sense of  a  converging
power series may be regarded as an unduly luxury. 
In this respect,  the above results may hold 
for   quaternions quantum mechanics at an intermediate energy  theory,  where
the combined symmetry includes  the  $SU(2)$ group. 
For   higher energies  we would expect the emergence of the  $SU(3)$ group
and  the  octonion algebra.

\end{document}